# Achieving Secure and Efficient Cloud Search Services: Cross-Lingual Multi-Keyword Rank Search over Encrypted Cloud Data

Xueyan Liu[1], Zhitao Guan[1], Longfei Wu[2], Zain Ul Abedin[1], Mohsen Guizani[3]

1. School of Control and Computer Engineering, North China Electric Power University, Beijing, China
2. Department of Mathematics and Computer Science, Fayetteville State University, Fayetteville, NC, USA
3. Electrical and Computer Engineering Department, University of Idaho, Moscow, ID USA

*Abstract*— Multi-user multi-keyword ranked search scheme in arbitrary language is a novel multi-keyword rank searchable encryption (MRSE) framework based on Paillier Cryptosystem with Threshold Decryption (PCTD). Compared to previous MRSE schemes constructed based on the k-nearest neighbor searchable encryption (KNN-SE) algorithm, it can mitigate some drawbacks and achieve better performance in terms of functionality and efficiency. Additionally, it does not require a predefined keyword set and support keywords in arbitrary languages. However, due to the pattern of exact matching of keywords in the new MRSE scheme, multilingual search is limited to each language and cannot be searched across languages. In this paper, we propose a cross-lingual multi-keyword rank search (CLRSE) scheme which eliminates the barrier of languages and achieves semantic extension with using the Open Multilingual Wordnet. Our CLRSE scheme also realizes intelligent and personalized search through flexible keyword and language preference settings. We evaluate the performance of our scheme in terms of security, functionality, precision and efficiency, via extensive experiments.

*Keywords—Cloud computing; searchable encryption; semantic search; privacy-preserving; top-k*

## I. INTRODUCTION

Today, cloud computing [1] has become a mature computing model which provides scalable and virtualized computing resources over the Internet. Cloud storage is a cloud computing system mainly used for data storage and management. Specifically, cloud storage is a new approach for providing storage resources on the cloud which brings great convenience for business and individual users[2-3].

The development of the cloud storage has been under the threats of various external and internal attacks. As a consequence, encryption technologies are commonly used to secure the data stored in cloud. However, this causes new challenges on the utility of cloud storage. To address the problem, searchable encryption [4] has been leveraged as an effective way to search over encrypted data. The searchable encryption technique has developed from single keyword search to multi-keyword ranked search.

Cao et al. proposed a widely used multi-keyword search framework MRSE [5] which supports for sorting results. By leveraging KNN-SE algorithm [6], the relevance score between the secure index and the query trapdoor is calculated through coordinate matching[7-8]. Yang et al. presented a novel multi-user multi-keyword rank search scheme [9] which addresses some problems in the previous MRSE schemes. By utilizing PCTD algorithm, the new MRSE [9] supports arbitrary language, flexible authorization, time-controlled revocation, flexible preference settings and does not require a global dictionary. It can also achieve simultaneous search within multiple users' documents using one single trapdoor. However, due to the exact match of keywords in their scheme, the multilingual search is limited to each language. It is impossible to search across languages. Moreover, it fails to satisfy the requirements for semantic search. Several papers (e.g., [10-14]) have studied related security issues.

To solve the above problems, we propose our **cross-lingual multi-keyword rank search** (CLRSE) scheme. Contributions of our work are as follows:

1) For the first time, we explore the problem of cross-lingual multi-keyword ranked search over encrypted cloud data. In contrast to [9], we build a cross-lingual target query upon Open Multilingual WordNet (OMW) [15] through language conversion and semantic extension to the original query, which can better eliminate the language barriers in searchale encryption.

2) Through flexible keyword and language preference settings, as well as automated calculation of preference scores for extended keywords according to the semantic , our scheme can achieve inteligent and personalized sorting search, and improve the accuracy of top-k search results.

3) We evaluate the performance of our scheme in terms of security, functionality, percision and efficiency through extensive experiments.

The rest of this paper is organized as follows. In section II, the problem formulation is given. In section III, our scheme is described in details. In section IV, the security analysis is stated. The performance of the proposed scheme is evaluated in Section V. In Section VI, the paper is concluded.

## II. PROBLEM FORMULATION

### A. System Model

The architecture of our scheme is shown in Fig. 1. There are six entities: Data Owner (DO) provides documents to be outsourced and generates the secure index for each document. Key Generation Center (KGC) takes a security parameter as input to generate all parameters and keys in the system. Cloud Platform (CP) has strong storage and computation capabilities, and stores user's files in the encrypted form. CP will also respond to users' computation and data retrieval requests. Computing Service Provider (CSP) is an online server with powerful computation capabilities. It executes interactive computations with CP. Data User (DU) generates a keyword trapdoor and issues data retrieval request to CP Query Transform Server (QTS) is a user-side server which converts user's original query to the target query.

### B. Design Goals

*1) Cross-Lingual Search:* Our CLRSE scheme should support cross-lingual search. The data user can issue queries in any language and specify the language type of the returned results.

*2) Query Semantic Extension:* The queries in our scheme are semantically extended before the trapdoor gene- ration to contain more semantical information.

*3) Personalized Ranked Search:* In our scheme, different users can specify scores for query keywords and languages according to their preferences. And the results will be sorted before returned to the data user.

*4) Security Goals:* We make the assumption that KGC is a fully trusted entity, CP and CSP are honest-but-curious, i.e., they are honest to execute the protocols but curious about users' data and will attempt to deduce any useful information during operations. But they will not be compromised simultaneously. The security goals of our system are defined as follows: *a). Confidentiality of outsourced data.* Our scheme should be able to protect the confidentiality of documents, indexes, language information, file IDs and sysmentric keys. CP and CSP should not deduce any useful information from the uploaded tuples and encrypted documents. *b). Unlinkability of query.* Our scheme should ensure that even if the data user issues two identical queries, they can be transformed into two completely different trapdoors. CP and CSP cannot deduce the relationship between trapdoors. *c). Privacy-Preserving.* CP and CSP cannot learn any userful infomation from the returned results.

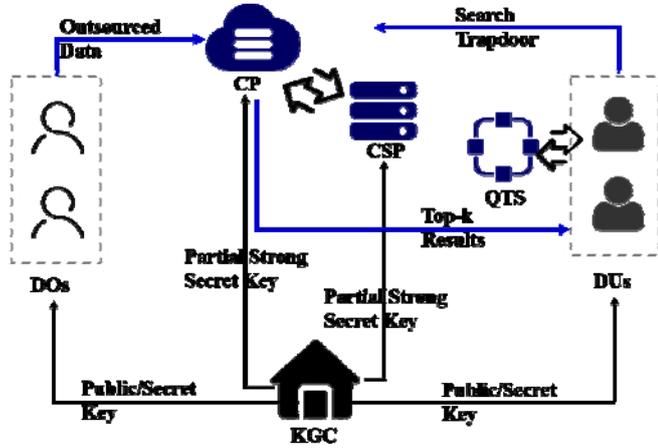

Fig. 1. System Model

## C. Notations

In table I, we list the notations used in CLRSE scheme.

TABLE I. NOTATIONS IN CLRSE

| Acronym | Descriptions |
|---|---|
| $pk_i$ | The public key of data owner i. |
| $pk_u$ | The public key of data user. |
| $sk_i$ | The secret key of data owner i. |
| $pk_\Sigma$ | The joint public key |
| $F_{i,j}$ | The j-th document of $A_i$ |
| $ID_{i,j}$ | The identifier of $F_{i,j}$ |
| $K_{i,j}$ | The symmetric key of $F_{i,j}$ |
| $I_{i,j}$ | The secure index of $F_{i,j}$ |
| $la_{i,j}$ | The language tag of $F_{i,j}$ |
| $\lambda$ | Strong secret key |
| $Td$ | Trapdoor |
| $Q$ | The data user's search query |

## D. Preliminaries

*1) PCTD (Paillier Cryptosystem with Threshold Decryption)*

PCTD is a public key cryptosystem with addition homomorphism and scalar multiplication homomorphism. The original double trapdoor public key cryptosystem (DT-PKC) [16] is a variant of Paillier's [17] and is constructed based on two difficult problems: the factorization of the modulus and solving discrete logarithm. Derived from DT-PKC, PCTD splits the strong secret key into shares and achieves partial decryption mechanism which makes it suitable for search protocols over encrypted data. Due to the distributed weak decryption, it has also been widely used in multi-key environments.

*2) OMW (Open Multilingual Wordnet)*

Open Multilingual Wordnet [15] offers an open network in a variety of languages, all of which are associated with the Princeton Wordnet of English (PWN). It intends to make wordnets in multiple languages easy to use. It is linked to Princeton WordNet 3.0 and its data has been extracted and normalized,. The Open Multilingual Wordnet and its components are open: anyone can freely use, modify, and share them for any purpose.

## III. CLRSE

The CLRSE scheme is described in detail in this section. It operates in six phases: Initialization, Documents Outsourcing, Cross-Lingual query construction, Trapdoor Generation, Top-k Rank Search.

## A. Initialization

In this phase, the system is initialized. KGC takes security parameter $l$ as input and randomly chooses two prime numbers $p$ and $q$ whose bit length is $l$. Then, KGC generates public parameter $(g, N)$ by randomly selecting $a \in Z^*_{N^2}$ and computing $N = pq$, $g = -a^{2N} \mod N^2$. The master secret key $\lambda$ is $lcm(p-1, q-1)$. KGC splits $\lambda$ into two parts $\lambda_1$ and $\lambda_2$ using the master secret key splitting algorithm [5]. They are sent to CP and CSP through a secure channel, respectively. Each user $i$ possesses a pair of private key $sk_i$ and the corresponding public key $pk_i = g^{sk_i}$. $sk_i$ is securely sent to user $i$. Finally, KGC generates a hash function $H: Z_N \to \kappa$ which can map an element in $Z_N$ to a symmetric key.

## B. Documents Outsourcing

**Step1** If data owner $i$ want to outsource his document $F_{i,j}$ to the cloud, he should construct a secure index for the file. Firstly, the data owner extracts a set of keywords $\{k_1, k_2, ..., k_n\}$ and assigns a tag $la$ such as 'eng' and 'cmn' which denotes the language information of the file. 'eng' and 'cmn' means that the document is in English and Chinese. In our scheme, we set the weight of the keyword as the product of $TF \times IDF$ (Term Frequency and Inverse Document Frequency), such that each value is an element in $Z_N$.

**Step2** Data owner $i$ executes the keyword to unique (K2U) bigInteger protocol described in **Algorithm 1**. K2U protocol maps each keyword of multi-language into a unique big integer $u \in Z_N$.

| **Algorithm 1**: Keyword to Unique BigInteger Protocol (K2U) |
|---|
| **Input :** keyword $k$ ( a string of length n) |
| **Output:** big integer $u$ |
| 1    Initialize $u = 0$ |
| 2    for $i = 0$ to $n-1$ do : |
| 3        $x[i]$ (hex) $\leftarrow$ unicode of $k[i]$ |
| 4        $y[i]$ (dec) $\leftarrow x[i]$ |
| 5        $s[i] \leftarrow y[i] \times 2^{16 \times i}$ |
| 6        $u \leftarrow u + s[i]$ |
| 7    Return $u$ |

As shown in **Algorithm 1**, each letter is represented by the hexadecimal Unicode and is converted to the decimal, then multiplied by the weight. The weight value of each letter is related to its location.

Next, each integer is encrypted by the PCTD algorithm. The ciphertext $[u]_{pk_i}$ encrypted with the owner's public key $pk_i$ is a binary group $(U_1, U_2)$, where $U_1 = pk_i^r (1+uN) \mod N^2$, $U_2 = g^r \mod N^2$ and $r$ is randomly chosen from $Z_{N^2}$. The secure index of file $F_{i,j}$ is constructed as follows:

$$I_{i,j} = \begin{cases} ([u_1]_{pk_i}, [\alpha_1]_{pk_i}) \\ ([u_2]_{pk_i}, [\alpha_2]_{pk_i}) \\ ... \\ ([u_n]_{pk_i}, [\alpha_n]_{pk_i}) \end{cases}$$

**Step3** Data owner $i$ chooses a number $K_{i,j} \in Z_N$ and encrypts $F_{i,j}$ to $[F_{i,j}]_{\kappa_{i,j}}$ using symmetric encryption. The symmetric key $\kappa_{i,j}$ is generated by $H(K_{i,j})$. Then, the data owner uploads the tuple $T_{i,j} == \{I_{i,j}, [ID_{i,j}]_{pk_i}, [K_{i,j}]_{pk_i}, [la_{i,j}]_{pk_i}\}$ together with the encrypted document to CP.

## C. Construction of Cross-Lingual query

If a data user want to make a search, he selects a set of query keywords $Q = \{qw_1, qw_2, ..., qw_m\}$ and assigns the preference scores of keywords as $P = \{\beta_1, \beta_2, ..., \beta_m\}$. Next, the user generates language tags $\lambda = \{lq_1, lq_2, ..., lq_t\}$ which indicate the type of file languages that the user expects to receive and assigns preference scores $P_l = \{\gamma_1, \gamma_2, ..., \gamma_t\}$ to each type of languages. Then, the data user sends tuple ($Q$, $P$, $\lambda$, $P_l$) to QTS. Upon reception of the tuple, QTS converts the original query into the target query through OMW. QTS constructs the cross-lingual target query in four steps: Language Conversion, Synonym Replacement, Reduction and Semantic Extension.

**Language Conversion and Synonym Replacement** After receiving ($Q$, $P$, $\lambda$, $P_l$), QTS locates the WordNets of the corresponding languages according to $\lambda$. For each keyword in $Q$, QTS search for it in OMW and replace each term with its synonym sets in the target languages. We set the preference score of each synonym as the product of the corresponding keyword score $\beta$ and the language score $\gamma$.

**Reduction** If there are two or more keywords in the original query that belong to a synonym set, there will be duplicate synonym sets in the extended query. The redundant synonyms will be deleted to reduce the overhead while improving the search efficiency. The simplified cross-lingual query contains several non-intersecting synonym sets.

**Semantic extension** In this step, we extend the simplified cross-lingual query semantically. In WordNet, if keywords are semantically related, they must have a certain kind of relationship such as synonym, hypernym/hyponym, meronym/holonomy. Through OMW, semantically related synonyms of each term are added into it. The similarity score of two terms $sim(t_1, t_2)$ can be easily obtained through OMW. For flexible control of the degree of semantic extension, we set a threshold $T(0 < T \leq 1)$, which is the minimum similarity score between existing synonym sets and extended semantic related keywords. ($T = 1$ means synonym replacement). Finally, we set the preference score of extended semantically related synonym $\eta$ as the product of the corresponding synonym score $\beta \cdot \gamma$ and the similarity score $sim(k, s)$.

## D. Trapdoor Generation

After the construction of cross-lingual query is completed, the original query is transformed from a collection of individual keywords into a collection of synonyms. To generate a trapdoor, the data owner encrypts each keyword in the target query and their corresponding preference scores. The trapdoor can be constructed as follows:

$$Td = \begin{cases} Td_1 \begin{cases} s_{1-lq_1}([w_1]_{pk_u}, [w_2]_{pk_u}, ...), [\beta_1 \cdot \gamma_1 \cdot 1]_{pk_u} \\ s'_{1-q_1}([w_1]_{pk_u}, [w_2]_{pk_u}, ...), [\beta_1 \cdot \gamma_1 \cdot sim(s_{1-q_1}, s'_{1-q_1})]_{pk_u} \cdots \\ ... \\ s_{m-lq_1}([w_1]_{pk_u}, [w_2]_{pk_u}, ...), [\beta_m \cdot \gamma_1 \cdot 1]_{pk_u} \\ s'_{m-q_1}([w_1]_{pk_u}, [w_2]_{pk_u}, ...), [\beta_m \cdot \gamma_1 \cdot sim(s_{m-q_1}, s'_{m-q_1})]_{pk_u} \cdots \\ [lq_1]_{pk_u} \end{cases} \\ ... \\ Td_t \begin{cases} s_{1-lq_t}([w_1]_{pk_u}, [w_2]_{pk_u}, ...), [\beta_1 \cdot \gamma_t \cdot 1]_{pk_u} \\ s'_{1-lq_t}([w_1]_{pk_u}, [w_2]_{pk_u}, ...), [\beta_1 \cdot \gamma_t \cdot sim(s_{1-q_t}, s'_{1-q_t})]_{pk_u} \cdots \\ ... \\ s_{m-lq_t}([w_1]_{pk_u}, [w_2]_{pk_u}, ...), [\beta_m \cdot \gamma_t \cdot 1]_{pk_u} \\ s'_{m-q_t}([w_1]_{pk_u}, [w_2]_{pk_u}, ...), [\beta_m \cdot \gamma_t \cdot sim(s_{m-q_t}, s'_{m-q_t})]_{pk_u} \cdots \\ [lq_t]_{pk_u} \end{cases} \end{cases}$$

### E. Top-k Rank Search

To issue a search request, the data user sends the trapdoor $Td$ and a parameter $k$ to CP. $k$ denotes the number of results the user wants to receive. In this phase, CP and CSP search over all documents in cloud for results that match the query trapdoor and pick out the $k$ results with the highest relevance scores. The following protocols in [9] are utilized in our proposed scheme:

Secure Addition Protocol across Domains (SAD):

$[A]_{pk_A} + [B]_{pk_B} \rightarrow [A+B]_{pk_\Sigma}$ $(pk_\Sigma = pk_A \cdot pk_B)$

Secure Multiplication Protocol across Domains (SMD):

$[A]_{pk_A} \cdot [B]_{pk_B} \rightarrow [A \cdot B]_{pk_\Sigma}$

Encrypted Keyword Equivalence Testing Protocol (KET): $[A]_{pk_A}, [B]_{pk_B} \rightarrow u^*$ ($u^*=1$ indicates that the two keywords are the same; $u^*=0$ otherwise.

Secure Top-k Data Retrieval Protocol across Domains (Top-k):

$(T_1, T_2, ..., T_n, k) \rightarrow (T_{max1}, T_{max2}, ...T_{maxk})$

**Step1** We apply the relevance score computation (RSC) protocol to calculate all relevance scores between search query and secure indexes. The details are described in **Algorithm 2**.

The RSC protocol takes a set of index tuples of all documents and the trapdoor as inputs, and outputs a collection of relevance score tuples. In **Algorithm 2**, we initialize an empty set $S$ to store the results.

*1)* In line 3-6, it utilizes the KET protocol to compare each language tag $[la]_{pk_i}$ in the document with all language tags $[lq]_{pk_u}$ in the trapdoor $Td$ and the returned result is $[d]_{pk_\Sigma}$. Then, CP and CSP jointly execute the paritital decryption algorithm to obtian $d$. If $d=1$, the corresponding file is the target document we are searching for; $d=0$, it is not.

*2)* In line 7-16, it determines the product of the weight score $\alpha_{ia}$ of keyword $k_{ia}$ and the preference score $\eta_{jt}$ of $w$.

If $k_{ia} = w$, the relevance score of $k_{ia}$ and $w$ is $[s_i]_{pk_\Sigma} = [\alpha_{ia} \cdot \eta_{jt}]_{pk_\Sigma}$ since

$[s_i]_{pk_\Sigma} = SMD$ ( $KET([k_{ia}]_{pk_i}, [w]_{pk_u})$, $[s']_{pk_\Sigma}$ )

$= SMD$ ( $[1]_{pk_\Sigma}$, $[\alpha_{ia} \cdot \eta_{jt}]_{pk_\Sigma}$ )

Otherwise, $[s_i]_{pk_\Sigma} = [0]_{pk_\Sigma}$ since

$[s_i]_{pk_\Sigma} = SMD$ ( $KET([k_{ia}]_{pk_i}, [w]_{pk_u})$, $[s']_{pk_\Sigma}$ )

$= SMD$ ( $[0]_{pk_\Sigma}$, $[\alpha_{ia} \cdot \eta_{jt}]_{pk_\Sigma}$ )

Then, the relevance score is added into I:

$[I'_i]_{pk_\Sigma} \leftarrow [I'_i]_{pk_\Sigma} \cdot [s_i]_{pk_\Sigma}$

*3)* In line 17, the tuple $S_i$ is added into $S$.

**Step2** After the execution of RSC protocol, $S$ stores all relevance score tuples that satisfy the data user's requirements

| **Algorithm 2**: Relevance Score Computation Protocol Across Domains (RSC) |
|---|
| **Input :** Index tuple set $<T_1, T_2, ..., T_n>$, **trapdoor** $Td$ |
| where $T_i = (I_i, [ID_i]_{pk_i}, [K_i]_{pk_i}, [la_i]_{pk_i})$ |
| $I_i = (I_{i_1}, I_{i_2}, ..., I_{i_{n_1}}), I_{ia} = ([k_{ia}]_{pk_i}, [\alpha_{ia}]_{pk_i})$ |
| $Td = (Td_1, Td_2, ..., Td_{n_2}), Td_j = (s_{j_1}, s_{j_2}, ..., s_{j_{n_3}}, [lq_j]_{pk_u})$ |
| $s_{j_t} = ([w_1]_{pk_u}, [w_2]_{pk_u}, ..., [w_k]_{pk_u}, [\eta]_{pk_u})$ |
| **Output:** relevance score tuple set $S = <S_1, S_2, ..., S_m>$ (m ≤ n) |
| 1   Initialize $S = \Phi$ |
| 2   for $i = 1$ to $n$ do |
| 3     Initialize $[d]_{pk_\Sigma} = [0]_{pk_\Sigma}$ |
| 4     for $j = 1$ to $n_2$ do |
| 5       $[d_j]_{pk_\Sigma} \leftarrow KET([la_i]_{pk_u}, [lq_j]_{pk_u})$ |
| 6       $[d]_{pk_\Sigma} \leftarrow SAD([d]_{pk_\Sigma}, [d_j]_{pk_\Sigma}$ |
| 7     CP and CSP jointly decrypts $[d]_{pk_\Sigma}$ |
| 8     if $d$ equals 1 |
| 9       for $j = 1$ to $n_2$ |
| 10        for $t=1$ to $n_3$ do |
| 11          for each $w$ in $s_{j_t}$ : |
| 12            CP and CSP jointly calculate |
| 13            $[u_i]_{pk_\Sigma} \leftarrow KET([k_{ia}]_{pki}, [w]_{pk_u})$ |
| 14            $[s'_i]_{pk_\Sigma} \leftarrow SMD([\alpha_{ia}]_{pk_i}, [\eta_{jt}]_{pk_u})$ |
| 15            $[s_i]_{pk_\Sigma} \leftarrow SMD([u_i]_{pk_\Sigma}, [s'_i]_{pk_\Sigma})$ |
| 16            $[I'_i]_{pk_\Sigma} \leftarrow [I'_i]_{pk_\Sigma} \cdot [s_i]_{pk_\Sigma}$ |
| 17        $S \leftarrow ([I'_i]_{pk_\Sigma}, [ID_i]_{pk_i}, [K_i]_{pk_i})$ |
| 18  Return $S$ |

for language. In this phase, CP and CSP take the collection of relevance score tuples $S$ and the parameter k as inputs, and executes the interactive privacy-preserving TOP-k protocol in [9]. The algorithm outputs k tuples $<T_{\max 1}, T_{\max 2}, ..., T_{\max n}>$ where $T_{\max i} = ([I'_{\max i}]_{pk_\Sigma}, [ID_{\max i}]_{pk_\Sigma}, [K_{\max i}]_{pk_\Sigma})$ which have the k highest relevance scores. Finally, the tuples are returned to the data user.

**Step 3** Upon receiving the tuple set, the data user firstly recovers every $ID_{\max i}$ and $K_{\max i}$ in the tuples by computing

$$ID_{\max i} = Dec([ID_{\max i}]_{pk_\Sigma}, sk_\Sigma)$$
$$K_{\max i} = Dec([ID_{\max i}]_{pk_\Sigma}, sk_\Sigma)$$
$$(Dec([m]_{pk_i}, sk_i) = L(C_1 / C_2^{sk_i} \mod N^2);$$
$$L(x) = (x-1)/N; [m]_{pk_i} = (C_1, C_2)$$

Then, the data user acquires each document's symmetric key by calculating $\kappa_{\max i} = H(K_{\max i})$. The user can obtain the encrypted documents secretly by utilizing the Private Information Retrieval (PIR) technology [18] which does not disclose the access pattern.

## IV. SECURITY ANALYSIS

### A. Protocol Security Proof.

Theorem 1. The proposed RSC protocol is secure to calculate the relevance scores on encrypted indexes and queries in the presence of semi-honest (non-colluding) attackers defined in [9].

Proof. Since SAD, SMD and KET are subprotocols of RSC which have been proved secure in [9] and the transmitted data is the ciphertext, the RSC protocol is also secure in the presence of the attackers.

### B. Security Goals Analysis

#### 1) Confidentiality of outsourced data.

Since the secure indexes are generated using PCTD, the file ID, the language tag and the symmetric key are also encrypted by PCTD, the confidentiality of outsourced tuples is based on the security of PCTD algorithm which have been proved in [9]. Since the documents are encrypted by symmetric keys generated by a hash function, we base the confidentiality of documents on the security of symmetric encryptions. Because the symmetric key is protected by PCTD, only data users who own the corresponding secret key can obtain the symmetric key to recover the documents.

#### 2) Unlinkability of Query

When generating a trapdoor, each keyword and the preference scores are separately encrypted using PCTD. During the execution of the PCTD encryption algorithm, random numbers are chosen to generate the ciphertext. Thus, even for the same plaintext, the ciphertexts generated by the PCTD encryption algorithm can be different. That is to say, even if the data user issues two identical queries, they can be transformed into two completely different trapdoors. CP and CSP cannot deduce the relationship between two trapdoors.

#### 3) Privacy-Preservation

In the query phase, the relevance scores are calculated by executing the RSC protocol. Since the security of the RSC protocol have been proved in Theorem1, the privacy of the relevance scores can be preserved. The top-k results are calculated by executing the TOP-k protocol which is proved secure in [9]. Besides, all transmitted data are ciphertexts. Hence, the privacy of the search algorithm is ensured. CP and CSP cannot learn any useful information from the obtained results.

## V. PERFORMANCE EVALUATION

In this section, we conduct extensive experiments to evaluate the performance of our proposed CLRSE search scheme. We implement our scheme on a PC with 3.40GHz Intel® Core™ i7-6700 processor and 16.0GB RAM. All cryptographic algorithms and protocols are implemented using Java language with the Eclipse IDE. For the implementation of the query extension, we access the open multilingual wordnet corpus through the (Python) Natural Language Tool-Kit wordnet interface (NLTK). Then, we construct a multilingual database which contains 2048 encrypted documents and their indexes in Chinese, English and other languages. We analyze the precision of our proposed search scheme and compare our scheme with the schemes in [19], [9] in terms of functionality and with MRSE in terms of efficiency.

### A. Functionality

TABLE II. FUNCTIONALITY COMPARISION

| | MRSE [5] | PRSE_1 [10] | PRSE_2 [10] | CLRSE |
|---|---|---|---|---|
| Multi-keyword Search | √ | √ | √ | √ |
| Top-k Ranked Search | √ | | √ | √ |
| Personalized Search | √ | √ | √ | √ |
| Non-predefined Keyword Dictionary | √ | | | √ |
| Multi-language Search | √ | | | |
| Semantic extension | | | √ | √ |
| Cross-Lingual Search | | | | √ |

### B. Precision

The precision of our proposed scheme indicates whether the top-k results meet the data users' needs. We conduct an investigation and randomly choose 100 users of different languages to test our system. We set the users' feedbacks in three levels: Satisfaction, Basic Satisfaction and Dissatisfaction. The survey results show that about three-fifths of the users are satisfied with the search results, and one-fifth of the users are basically satisfied.

### C. Efficiency

#### 1) Document Outsourcing

In our proposed scheme, the outsourcing of documents mainly involves three steps: keyword extraction and weight value calculation, security index generation, index tuple generation. Since the first step can be completed ahead of time, the execution time of document outsourcing in our experiment is the sum of the execution time of the other two steps.

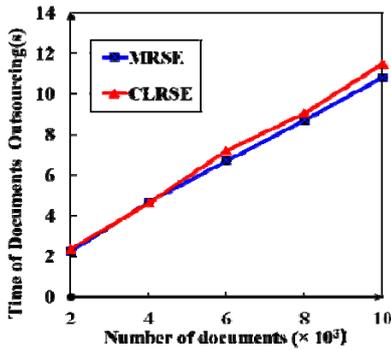 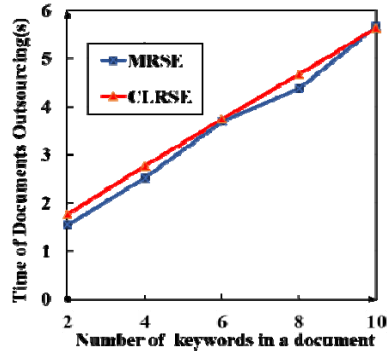 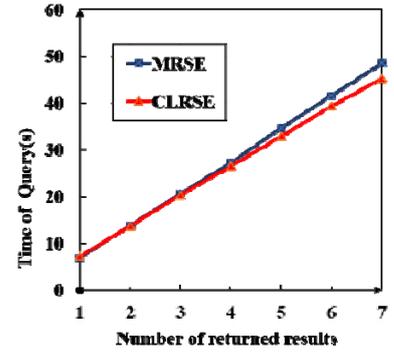

Fig. 2. Time of documents outsourcing for the different number of documents, the number of keywords in documents is 8.

Fig. 3. Time of documents outsourcing for the different number of keywords in a document, the number of documents is 4096.

Fig. 4. Response time: For the different retrieved number of documents (k), the number of keywords in original query is 4; the threshold T is 0.5.

We compare the document outsourcing time of our CLRSE scheme with the MRSE scheme. Since the execution time of all algorithms and protocols are affected by the bitlength of $N$, we set the bitlength of $N$ in our scheme as 1024 to achieve 80-bit security level [20]. The secure index generation phase of the two schemes is roughly the same. The total running time of this phase is proportional to the number of documents and the number of keywords in each document. In the third phase, our scheme requires an additional encryption of language tags which has only a slight impact on the document outsourcing efficiency. Fig.2. and Fig.3. show that the document outsourcing time of the two schemes are basically linear with the size of document collection and the number of keywords in a document. The document outsourcing efficiency of CLRSE is roughly equivalent to that of MRSE.

*2) Query*

The query response time is defined to measure the query efficiency, which refers to the period from the issue of the query request to the reception of the search results. For example, the data user specifies four Chinese keywords as the original query and attempts to get Chinese and English search results with T set to 0.5. Our CLRSE scheme involves an additional operation of cross-lingual query construction before trapdoor generation. Fig.4. shows that the query response time of the two schemes grows linearly with the number of the retrieved documents. When k=1, the time cost of our scheme is slightly longer than MRSE. However, the time cost of MRSE rapidly increases as the number of retrieved documents becomes larger. Therefore, when the number of returned results required by the user is large, the search efficiency of our scheme is significantly higher than that of MRSE.

Compared to MRSE, our scheme contains an extra process of query extension. The time overhead of this part is negligible. All the time-consuming operations in both schemes happen in the trapdoor generation phase and the top-k rank search phase. In our CLRSE scheme, the size of keyword collection in query is larger due to the extension of the original query, which will cause an increase of the trapdoor generation overhead. On the other hand, our scheme filters out the document collection of the target languages by language matching which can significantly reduce the runtime of the top-k protocol. In addition, all protocols running on the CP and CSP can be calculated in parallel which can greatly improve the efficiency of the protocol execution.

## VI. CONCLUSION

In this paper, we investigated the problem of cross-lingual multi-keyword rank search with semantic extension over encrypted data and proposed a cross-lingual multi-keyword rank search scheme CLRSE using OMW. Our proposed scheme made valuable contribution to searchable encryption in that it further eliminates the language barriers in searchable encryption. Additionally, our scheme implemented the flexible keyword and language preference settings, as well as the automated calculation of preference. The performance of our scheme is evaluated and compared with previous schemes. In the future, we will focus on the improvement of the top-k search efficiency in our scheme.

ACKNOWLEDGMENT

This work is supported by Beijing Natural Science Foundation under grant 4182060.